# Towards Efficiently Solving Quantum Traveling Salesman Problem

Debabrata Goswami, Harish Karnick, Prateek Jain, and Hemanta K. Maji
Department of Computer Science and Engineering
Indian Institute of Technology, Kanpur-208 016
(Dated: November 2, 2004)

We present a framework for efficiently solving Approximate Traveling Salesman Problem (Approximate TSP) for Quantum Computing Models. Existing representations of TSP introduce extra states which do not correspond to any permutation. We present an efficient and intuitive encoding for TSP in quantum computing paradigm. Using this representation and assuming a Gaussian distribution on tour-lengths, we give an algorithm to solve Approximate TSP (Euclidean) within **BQP** resource bounds. Generalizing this strategy for any distribution, we present an oracle based Quantum Algorithm to solve Approximate TSP. We present a realization of the oracle in the quantum counterpart of **PP**.

03.67.Lx

*Introduction*—Quadratic speed up achieved by Grover's[1] search algorithm for unstructured database is optimal[2,3,4]. Direct attempts to solve **NP**-complete problems are bound to fail unless some structure is identified in the search space. Exponential speedup in many problems is achieved by exploiting some underlying property of the search space, for example, Shor's Factorization Algorithm[5], Deutsch and Jozsa's algorithm[6], Childs *et. al.*[7] random walk algorithm.

Any search algorithm to solve Approximate Traveling Salesman Problem[8] (Approximate TSP) is bound to fail unless it uses some additional information about the search space. The best known classical algorithms for Euclidean Approximate TSP are Christofides' algorithm[9] and PTAS[10]. Farhi *et. al.*[11,12] proposed a quantum optimization technique for Satisfiability problem using adiabatic evolution, but this method is not able to guarantee bounded error in polynomial time. Hogg *et. al.*[13,14,15] have proposed a quantum optimization technique using mixing operators. They have simulated it for Satisfiability Problem and Approximate TSP. The method performs well on average, but a bounded error in polynomial time cannot be guaranteed by this algorithm. Hogg *et. al.* also propose an encoding that introduces states which do not correspond to any permutation. So, efficient representation of TSP is another issue that needs to be addressed.

We propose a new encoding scheme for permutations which can be recursively generated. Using this representation scheme and assuming a *Gaussian distribution* on the tour lengths we present a **BQP** algorithm to solve Approximate TSP (Euclidian). We propose another algorithm to solve Approximate TSP using an oracle which answers queries about tour-length distribution. This algorithm gives correct answers with high probability but is not in **BQP**. However, this algorithm has the advantage that it does not assume any distribution on the tour lengths.

*Encoding Scheme for TSP*—Given a graph on $n$ vertices, every permutation of the set $\{1,2,\ldots,n\}$ defines a possible Hamiltonian cycle over the graph. So, if we have an entanglement of all possible permutations, information about whether a permutation defines a Hamiltonian cycle over the graph or not can be associated with it. If a permutation represents a valid Hamiltonian cycle then we associate its tour length with the permutation. We establish a bijection between the set of all permutations of $\{1,2,...,n\}$ and the set $\{(a_1,a_2,...,a_n) \mid 1 \leq a_i \leq i \ \forall i 1 \leq i \leq n\}$, which we will call the *encoding set*. We give an inductive definition of this mapping function. For the base case of $n=1$, the function maps $(a_1 = 1)$ to the permutation 1. For $n>1$, it maps $(a_1,a_2,...,a_n)$ in the following manner:

1. From the elements $(a_1,a_2,...,a_{n-1})$, create the corresponding permutation of the set $\{1,2,...,n-1\}$.
2. Insert $n$ between the $(a_n-1)$-th and $a_n$-th elements of the previous permutation (for $a_n = 1$ we insert in the beginning and for $a_n = n$ we insert it at the end).

A simple $O(n^2)$ classical algorithm can be designed which implements this mapping. So, there exists a quantum algorithm to implement this mapping in polynomial time (using synchronization lemma[16]).

Note that if the *i*-th elements differ in two encodings then *i* is inserted at different positions and subsequent permutations which are generated are different. Therefore, this function is one-one. Moreover, size of the range and domain are equal ($= n!$). Thus, the function is a bijection and it suffices to obtain an entanglement of all elements in the encoding set.

*Circuit to generate entanglement of all Permutations' encodings*—In order to generate entanglement of all possible encodings, we use $n$ sets of registers $R_1,...R_n$ to store information related to the entries $(a_1,...,a_n)$. The $i$-th register ($R_i$) has $i+1$ qubits. $R_{i,j}$ represents the $j$-th qubit of $i$-th register. At any given time, exactly one of the qubits in a register is in state $|1\rangle$ and rest are in state $|0\rangle$. If $R_{i,j} = |1\rangle$ and other qubits in $i$-th register are in $|0\rangle$ state, then we write $R_i = |j-1\rangle$ (this means that $a_i = j-1$). Our initial state is:

$$\Phi_1 = |1\rangle \overbrace{|0\rangle...|0\rangle}^{(n-t)} \quad \text{(base condition of the function)}$$

After $t$ iterations we will have the state (ignoring the normalization constant):

$$|\Phi_t\rangle = \left(\sum_{a_1=1}^{1}|a_1\rangle\right) \otimes ... \otimes \left(\sum_{a_t=t}|a_t\rangle\right) \overbrace{|0\rangle...|0\rangle}^{(n-t)}$$

$$= \sum_{a_1=1}^{1}...\sum_{a_t=1}^{t}|a_1\rangle...|a_t\rangle \overbrace{|0\rangle...|0\rangle}^{(n-t)}$$

We apply the gate $G_{t+1}$ to obtain $\Phi_{t+1}$ from $\Phi_t$. Gate $G_{t+1}$ can be described by:

Triggered by: $\quad R_{t,1} = |0\rangle$

$\quad\quad\quad\quad\quad\quad R_{t+1,1} = |1\rangle$

Effects the change: $\quad R_{t+1,j} \to \neg R_{t+1,j} \quad \forall j 1 \leq j \leq t+1$

where $\neg R_{t+1,j}$ represents complement of $R_{t+1,j}$. $R_{t,1} = |0\rangle$ implies that the values of $a_t$ have been set and $R_{t+1,1} = |1\rangle$ implies that values of $a_{t+1}$ are not yet decided. So, this

step changes $R_{t+1}$ from $|0\rangle$ to the state $\sum_{a_{t+1}=1}^{t+1}|a_{t+1}\rangle$ in the $(t+1)$-th iteration. At the end of $n$ iterations, we obtain the state representing the entanglement of all possible permutation encodings:

$$|\Phi_n\rangle = \left(\sum_{a_1=1}^{1}|a_1\rangle\right) \otimes \ldots \otimes \left(\sum_{a_n=1}^{n}|a_n\rangle\right)$$

$$= \sum_{a_1=1}^{1}\ldots\sum_{a_n=1}^{n}|a_1\rangle\ldots|a_n\rangle$$

We modify the previous set of gates $\mathcal{G}$ to obtain gates which result in a final wave where result in a final wave where the probability of observing a particular permutation is proportional to $\alpha^{\text{-tour length}}$, $\alpha > 1$. The $\mathcal{G}$ gates transformed a permutation of $t$ points into an equal superposition of all possible permutations resulting from inserting the $(t+1)$-th point. Given a permutation of $t$ points, we know the exact increment in each tour due to the insertion of the $(t+1)$-th point at each of $t+1$ possible positions (in all cases the length increases because the space is Euclidean). So, we can redistribute the probabilities in the resulting $t+1$ states such that they are proportional to $\alpha^{-\text{increment}}$. These gates, after $n$ iterations, result in a superposition of the encoding set, such that the probability of observing a particular encoding is proportional to $\alpha^{\text{-tour length}}$. So, we obtain the wave: $\sum_{\tau \text{ is a permutation}} a^{-\tau_{\text{length}}}|\tau\rangle$.

*Algorithm for solving Approximate TSP assuming a distribution*—Given a graph, with a Euclidean norm, we scale it to fit within the smallest axis parallel square. Since there is no smaller square which can contain the graph completely, there must be at least two vertices on opposite sides of the square. We now scale the graph such that this square is a unit square. Observe the following two properties for such graphs in unit squares:

1. Tour lengths are $O(n)$: Maximum edge length in a unit square is $\leq \sqrt{2}$, so the maximum tour-length that is possible in a normalized graph is $\leq \sqrt{2}n$. This bound is tight and nothing better can be obtained (consider $n = 2k$ points in a unit square with all vertices lying on one vertex of the square and all even vertices lying on the diagonally opposite vertex of the square. Then any tour which alternately chooses an odd numbered and even numbered vertex is of length $\sqrt{2}n$).

2. Tour lengths are $\Omega(1)$: Without loss of generality, assume that the two vertices which lie on opposite sides of the square are on edges which are parallel to the X-axis. Consider any tour, and project its edges on the Y-axis. Edge lengths are greater than their projections, so length of any tour is greater than its projection on the Y-axis. The projected tour has length 2, s all tours have tour length $\geq 2$. One can also verify that that this bound is tight (consider $n$ points lying on a straight line parallel to the Y-axis and the X-coordinates of the vertices of a tour form a bitonic sequence, then it has length 2).

Consider a fully connected Euclidean TSP instance. We assume that the tour-length distribution is Gaussian with the hump of the distribution lying between the minimum and maximum tour-lengths. For this TSP instance we prepare the wave

$$|\Psi_\alpha\rangle = \sum_{\tau \text{ is a permutation}} \alpha^{-\tau_{\text{length}}} |\tau\rangle$$

Given an $\varepsilon$ close to 0, we would like to choose $\alpha$ such that the probability of observing a tour with tour-length (by reading $|\Psi_\alpha\rangle$) within $(1+\varepsilon)$ of optimal tour-length is greater than a non-zero constant. Assume $g(x)$ is a Gaussian centered at some point in the range $[x_{min}, x_{max}]$. $x_{min}$ is minimum tour-length and $x_{max}$ is maximum tour length. $\alpha$ is a suitable parameter to be decided in the prepared wave. The probability of observing a tour with tour-length at most $(1+\varepsilon)$ times the optimal tour length is a function of $\alpha$ and could be expressed as:

$$\sigma = \frac{\int_{x_{min}}^{x_{min}(1+\varepsilon)} \alpha^{-x} g(x) dx}{\int_{x_{min}}^{x_{max}} \alpha^{-x} g(x) dx}$$

We will make two assumptions regarding the nature of $\sigma$ (the variance of the Gaussian). We will assume that $\sigma$ grows slower than a polynomial in $n$ and $1/\sigma$ grows slower than a polynomial in $n$. Now, if we substitute a new variable for $x/\sigma$, then we get the following properties in the new coordinate system:

1. $\sigma \leq p'(n)$ eventually: So $x_{min} \geq \dfrac{2}{p'(n)} \Rightarrow x_{min} \geq \dfrac{1}{p(n)}$ in new coordinate system.

2. $\dfrac{1}{\sigma} \leq q'(n)$ eventually: So $x_{max} \leq \sqrt{2} q'(n) \Rightarrow x_{max} \leq q(n)$ in new coordinate system.

Henceforth, we will assume that $\sigma = 1$ and prove our results with the above constraints.

Now, observe that $\alpha^{-x} = e^{-(\ln a)x} = e^{-2kx}$, where $2k = \ln a > 0$. The function that we intend to integrate becomes $e^{-2kx} e^{-x^2} = e^{k^2} e^{-(x+k)^2}$. Multiplying by $\alpha^{-x}$ gives another Gaussian with center shifted to the left by $k > 0$. As $\alpha$ varies, $k$ varies with it and the resulting set of functions is the set of all Gaussians shifted left from the original one. Now, we reduce the problem to analyzing the following function:

$$h(x) = \frac{\int_{x}^{x+\varepsilon x_{min}} e^{-x^2} dx}{\int_{x}^{x+(x_{max}-x_{min})} e^{-x^2} dx}$$

We are justified in doing so, because $e^{-2kx}$ serves to shift the center of $g(x)$ to the left by $k$. This new center is taken as 0 and $x_{min}$ goes to $x_{min} - k$ in the new co-ordinate frame, which is taken as the variable $x$ to account for variable $k$ in the original problem.

Now, we set $x = 0$, i.e. we choose $k$ such that the center of the shifted Gaussian is at $x_{min}$ of the original problem. So, the shift is $O(q(n))$. Therefore, $\alpha = e^{2k} = e^{O(q(n))}$. We recall the two inequalities $x_{min} \geq 1/p(n)$ and $x_{max} \leq q(n)$ (where $p(n)$ and $q(n)$ are fixed polynomials). Now, we see:

$$h(0) > \frac{\int_0^{0+\varepsilon x_{min}} e^{-x^2} dx}{\int_0^{x_{max}} e^{-x^2} dx} \geq \frac{\int_0^{\frac{\varepsilon}{p(n)}} e^{-x^2} dx}{\int_0^{q(n)} e^{-x^2} dx}$$

$$> \sqrt{\frac{1-e^{-\frac{\varepsilon^2}{p(n)^2}}}{1-e^{-2q(n)^2}}} \geq \frac{\varepsilon^2}{p(n)^2}$$

$$\text{when } \varepsilon \to 0 \left( \because x < \frac{1}{4} \Rightarrow x \geq 1 - e^{-x} \geq x^2 \right)$$

---

**Algorithm 1:** **BQP** algorithm for Approximate TSP (Euclidian) assuming Gaussian distribution on the tour-lengths

1. Input a graph over $n$ vertices and $\varepsilon$
2. Set $\alpha = e^{O(q(n))}$, and prepare $|\Psi_\alpha\rangle$
3. Read the wave and find out tour-length of the observed tour
4. Repeat Step 2 and 3 $O\left(\frac{p(n)^2}{\varepsilon^2}\right)$ times
5. Return the tour with minimum tour-length

---

Prepare the wave $|\Psi_\alpha\rangle$ with $\alpha = O(q(n))$ (as described in last section) and read it to find a tour. If we repeat this procedure $O\left(\frac{p(n)^2 r(n)}{\varepsilon^2}\right)$ times ($r(n)$ is a polynomial in $n$) with $\alpha$ set as $e^{O(q(n))}$, then the probability that we do not get any tour in $\varepsilon$ neighborhood of the optimal tour-length in $\gamma \frac{p(n)^2 r(n)}{\varepsilon^2}$ repetitions of the experiment is: $\left(1 - \frac{\varepsilon^2}{p(n)^2}\right)^{\gamma \frac{p(n)^2 r(n)}{\varepsilon^2}} < e^{-\gamma r(n)}$.

Therefore, we get a tour in an $\varepsilon$ neighborhood of the optimal tour with a probability greater than $1 - \frac{1}{e^{\gamma r(n)}}$. Hence, in time which is polynomial in $n$ and $\frac{1}{\varepsilon}$ we get a tour which within $(1+\varepsilon)$ times the optimal tour with very high probability.

*Oracle Algorithm and implementation*—In the previous section, the assumption that the tours of a randomly generated Euclidean TSP instance in a unit square have a Gaussian distribution helped us to solve the problem efficiently. Now, we consider the question of whether we can solve TSP efficiently if we have an oracle that can give us information about the possible distribution of tour lengths in a particular TSP instance. Assuming an oracle exists that can tell whether there are any tours with lengths in the range $[a,b)$ we give an efficient algorithm to solve Approximate TSP.

Assume that we have an equal superposition of all possible permutation of the set $\{1,2,...,n\}$. We know that given a permutation it is possible for a classical Turing Machine to verify in polynomial time whether that permutation is valid and has a tour length in the range $[a,b)$ *i.e.*, the tour length is actually in the range $[a,b+\delta)$. So, there exists a quantum Turing Machine which runs in polynomial time and solves this problem. We dovetail this machine and the circuit that produces an equal superposition of all permutations. This composite machine $M_1$ puts a validity qubit in state $|1\rangle$ if it is a valid tour with desired tour length, otherwise in state $|0\rangle$. We create another machine $M_2$ which puts its qubit in the state $\frac{|0\rangle+|1\rangle}{2}$ if it is a valid tour with desired tour length else puts it in the state $\frac{|0\rangle-|1\rangle}{2}$. We now consider the machine $M$ as described in the following algorithm:

---
**Algorithm 2: PP oracle realization**

1. Create two independent equal superpositions of all possible permutations.
2. Apply $M_1$ to the first wave and apply $M_2$ to the second wave.
3. Read the two validity bits of these machines using the measurement operators $\{|0\rangle\langle 0|, |1\rangle\langle 1|\}$. Allow other qubits to decohere.
4. If both are 0 return *false* else *true*.

---

Assume that there are $m$ valid tours out of a total of $N = n!$ tours. Probability that both qubits on reading give 0 is $\left(1-\frac{m}{N}\right) \times \frac{\left(\sqrt{1-\frac{m}{N}} - \sqrt{\frac{m}{N}}\right)^2}{2}$. If there are no valid tours, then $M$ returns *false* with probability $\frac{1}{2}$ else returns *true* with probability $> \frac{1}{2}$. Here with high probability the answer given by $M$ is correct, but this probability can not be directly amplified as the gap between the probabilities (when $m=0$ and $m=1$) decreases faster than inverse of any polynomial in $n$. So, this oracle is in the quantum counterpart of classical **PP**.

We divide the range $[2, \sqrt{2}n]$ into $\varepsilon$ size ranges and number them serially from 1 to $\left[\frac{\sqrt{2}n-2}{\varepsilon}\right]$. We search linearly with $\delta = \varepsilon$ on these ranges and try to find out the first range that has a valid tour in it. Let $i_0$ be the first range which has a valid tour such that its tour-length is in its range. We start with an equal superposition of all permutations. We dovetail a machine that finds the range to which a tour length belongs. This results in an equal superposition of all permutations and its corresponding range number in a separate register. Then as a final step, we project the $i_0$-th range and the decohered bits of the wave give us a permutation.

Since no bin with index less than $i_0$ has a tour, the optimal tour also lies in the $i_0$-th bin. The tour observed in the previous algorithm has length at most $\varepsilon + \delta = 2\varepsilon$ greater than the optimal tour. We know that every tour has tour length greater than $2$. So, we get the result that the obtained tour's length is less than $(1+\varepsilon)$ times the optimal tour-length. So, the obtained tour is an $(1+\varepsilon)$ approximation of the optimal tour.

Each iteration uses a call to the oracle and has $O(1)$ time requirement and there are $O\left(\frac{n}{\varepsilon}\right)$ iterations. So, we get an algorithm in the class $P^X$, where X is the class in which oracle lies. Our algorithm gives correct answers with high probability, however, a bounded error in probability cannot be guaranteed in polynomial time. When interpreted in classical complexity theory, the result is in lines of Toda's theorem, but an interesting fact is that we can try to use more than 2 different bases (here we used $\{|0\rangle, |1\rangle\}$ and $\left\{\frac{|0\rangle+|1\rangle}{\sqrt{2}}, \frac{|0\rangle-|1\rangle}{\sqrt{2}}\right\}$ bases sets in the oracle). This may not be possible in the standard Quantum Turing Machine Model, but there may be a bounded error algorithm in polynomial time for other quantum computation models.

---

**Algorithm 3:** General Algorithm to solve TSP

1. Divide the range $[2, \sqrt{2}n]$ into ranges of size $\varepsilon$ and number them from 1 to $\left[\frac{\sqrt{2}n-2}{\varepsilon}\right]$.
2. Set $\delta = \varepsilon$
3. Sequentially for each range from 1 to $\left[\frac{\sqrt{2}n-2}{2\varepsilon}\right]$ query the oracle with its search range.
4. If $i$ is the first index for which oracle returns *true* then set $i_0 = i$.
5. Create the wave $\sum_{\tau \text{ is a permutation}} |\tau\rangle \#|\tau\rangle$, $\#\tau$ is the range number in which $\tau$'s tour-length lies.
6. Project the qubits storing range information on $|i_0\rangle\langle i_0|$ and let the qubits storing the tour-length information decohere.
7. Return the $\tau$ obtained in the qubits storing the permutation information.

---

*Conclusion*—To the best of our knowledge, there is no quantum or classical algorithm which guarantees bounded error performance in polynomial time for any generic class of Traveling Salesman Problem. The Letter shows that if we assume a Gaussian distribution on the tour-lengths of all possible Hamiltonian cycles, then we can solve Approximate TSP in **BQP** resource bounds. Exact distribution of tour lengths may not be known. Oracle algorithm presents a method where we use an oracle to answer simple queries about Hamiltonian cycles' properties. We present a **PP** algorithm to realize an oracle which provides sufficient information to help solve Approximate TSP. Although this means that the algorithm does not guarantee bounded error in polynomial time, but it gives correct answer with high probability.

The results presented here can be considered amongst the few optimistic results on TSP. The methodology presented here provides a general framework within which one can use better oracles to obtain performance enhancement. There are a couple of evident extensions of the work presented in this Letter. We can analyze the effect of using multiple bases instead of two bases used in the oracle circuit presented in this Letter. Otherwise one can study oracle realizations in other models of quantum computations. If there are other models of quantum computation in which we can efficiently solve Approximate TSP with bounded error, then **NP** problems could be efficiently solved in those models of quantum computation.

---